\documentclass[conference]{IEEEtran}

\usepackage{algorithm}
\usepackage{algorithmic}
\usepackage[font=small]{caption}
\usepackage[labelsep=period]{caption}
\usepackage{float}
\usepackage[latin1]{inputenc}
\usepackage{graphicx}
\usepackage[cmex10]{amsmath}
\usepackage{cite}
\usepackage{amsfonts}
\usepackage{amsmath}
\usepackage{amsthm}
\usepackage{mathrsfs,dsfont}
\usepackage{amssymb,mathrsfs}
\usepackage[mathscr]{euscript}
\usepackage{epstopdf} 
\usepackage{color}
\usepackage{pgfplots}
\pgfplotsset{compat=newest}
\usetikzlibrary{plotmarks}
\usepackage{grffile}
\usepackage{amsmath}
\usepackage{caption}
\usepackage{subcaption}
\usepackage{balance}

\newcommand{\argmax}{\arg\!\max}

\title{ 
{Mobile Cellular-Connected UAVs: \\ Reinforcement Learning for Sky Limits} 
}
\author{  M. Mahdi Azari
, Atefeh Hajijamali Arani, Fernando Rosas 
}

\begin{document}

\maketitle

\begin{abstract}
A cellular-connected unmanned aerial vehicle (UAV) faces several key challenges concerning connectivity and energy efficiency. Through a learning-based strategy, we propose a general novel multi-armed bandit (MAB) algorithm to reduce disconnectivity time, handover rate, and energy consumption of UAV by taking into account its time of task completion. By formulating the problem as a function of UAV's velocity, we show how each of these performance indicators (PIs) is improved by adopting a proper range of corresponding learning parameter, e.g. 50\% reduction in HO rate as compared to a blind strategy. However, results reveal that the optimal combination of the learning parameters depends critically on any specific application and the weights of PIs on the final objective function.
\end{abstract}


\section{Introduction}


Unmanned aerial vehicles (UAVs)--so-called drones--have many beneficial applications such as environmental sensing, monitoring, and telecommunications 
\cite{mozaffari2019tutorial}. 
Critical for these applications is communication technology that can ensure UAV's connectivity, allowing a safe, reliable, and secure use of drones. In particular, a reliable beyond visual line-of-sight (BVLoS) command and control (C\&C) of UAVs is instrumental for enabling UAVs autonomous operation. 
Cellular connectivity has been proposed as a suitable candidate to serve UAVs' needs, and the development of this is an active endeavour in academia and industry \cite{zeng2019accessing,azari2019cellular}.

Although the use of cellular networks for UAVs communication seems to often be a win-win situation for both cellular and UAV operators \cite{azari2019cellular}, several challenges need to be addressed before largely launching such an idea. 
One important challenge is to provide adequate connectivity time to UAVs with respect to ground terminals, which can be particularly demanding for UAVs flying at high altitudes due to interference \cite{azari2019cellular}. On the other hand, a highly mobile UAV may frequently change its serving cell, and therefore it may trigger subsequent HOs. 
Since a HO procedure requires additional signaling overhead which causes  service interruptions, 
a high rate of HOs deteriorates the communication link reliability. Accordingly, the UAV's HO rate needs to be managed properly along its trajectory. Furthermore, UAVs in general are battery-limited, which restricts their operational lifetime \cite{zeng2019accessing}. As the energy consumption of UAV remarkably depends on its velocity, the speed of UAV needs to be optimized. Furthermore, the speed of UAV considerably influences
the connectivity time and HO rate. 
All aforementioned factors need to be taken into account when designing UAV operation, making this a highly non-trivial task.



The performance of cellular-connected UAVs and solutions to improve connectivity time and reliability of the corresponding link have been mostly addressed using model-based approaches 
\cite{azari2019cellular,azari2019uav,amer2020mobility}. In \cite{azari2019cellular,azari2019uav}, the authors thoroughly studied the coverage and rate performance of cellular networks for UAVs with and without HO effects. Also, HO analysis of cellular-connected UAVs has been investigated in 
\cite{amer2020mobility,fakhreddine2019handover}. 
The UAVs energy consumption is discussed in \cite{zeng2019accessing}, and the optimal trajectory design of UAVs by considering the energy consumption is studied in \cite{sallouha2018energy}. Besides, there is a growing interest in applying machine learning into UAVs communication and networking  
\cite{hu2020reinforcement,arani2020learning}. A few recent studies adopted different machine learning (ML) approaches to alleviate the detrimental effect of HOs \cite{azari2020machine,chen2019efficient,chowdhury2020mobility}. However, the existing works do not concretely take into account the stringent rate requirement of UAVs in the learning process, which makes unclear whether the proposed solutions are capable of delivering the target rate and connectivity time. 
Moreover, to our best knowledge, the limited energy capacity of UAVs for their task accomplishments is neglected in ML literature on UAVs.  

In this paper, we introduce a novel learning-based strategy to manage the mobility of UAVs for connectivity and reliability by taking into account the UAV's energy consumption and time of task completion. We first formulate the problem as a function of UAV's velocity, and then propose the sage of reinforcement learning (RL) to solve the problem. We propose a novel algorithm based on multi-armed bandit (MAB) to dynamically adjust the speed of UAV. We show that our proposed strategy significantly decreases HO rate, disconnectivity time, energy consumption, and time of task completion. Moreover, depending on the wight of each metric, by adopting proper values for the learning parameters, we observe an important overall performance 
 improvement with respect to a benchmark. 

The rest of this paper is organized as follows. In Section \ref{sec:system_modeling}, we present the network model and considered metrics. In Section \ref{sec:problem_statement}, the preliminaries are elaborated, and the problem is formulated. We propose our learning-based solution in Section \ref{sec:ML_Solution}, and in Section \ref{sec:numerical_results}, numerical results are presented. Finally, the paper is concluded in Section \ref{sec:conclusion}. 


\section{System Modeling} \label{sec:system_modeling}

In this section, we introduce the system model, which includes the network topology (Section \ref{subsec:topology}) and the channel propagation features (Section \ref{subsec:channel}), and the key performance indicators (KPIs) in Section \ref{subsec:KPI}.

\subsection{Network Topology}\label{subsec:topology}

\begin{figure}[t]
\centering
\includegraphics[width=0.9\columnwidth]{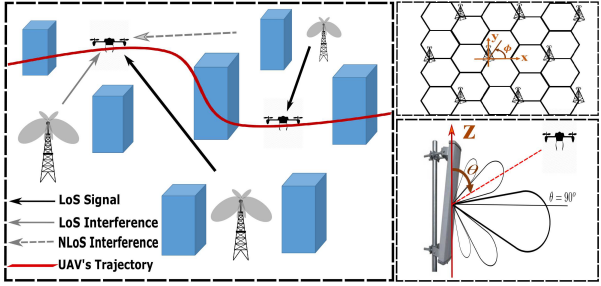}
	\caption{3D network representation of a mobile cellular-connected UAV.}
	\label{SystemModel}
	\vspace{-5mm}
\end{figure}

We consider the downlink of a wireless cellular network carried out by a set of base stations (BSs) deployed on a hexagonal layout as shown in Fig. \ref{SystemModel}. We assume each site consists of three co-located BSs of height $\mathrm{h_b}$,  and each BS covers  an angular interval of $120^o$ in horizontal 2D plane. 
We assume that each BS has an active user per time and frequency, which generates inter-cell interference to neighboring  cells working at the same frequency band. Each BS transmits with power of $\mathrm{P_b}$ distributed over the corresponding bandwidth. Therefore, the transmit power over each physical resource block (PRB) is equal to $\mathrm{P_b}/\mathrm{n_t}$, where $\mathrm{n_t}$ represents the total number of available PRBs.

Within this setup, we consider a UAV flying at altitude $\mathrm{h_u}$, which requires cellular connectivity. The  UAV is associated with the BS that provides the strongest signal strength. We further assume that the network allocates $\mathrm{n_u}$ PRBs for the UAV-BS link.


The trajectory of the UAV, denoted by $\mathbf{q_{tr}}$, might be regularly updated during its mission by the cellular controller. 
The UAV flies with velocity $V \in [\mathrm{V_{min}},\mathrm{V_{max}}]$ 
where $\mathrm{V_{min}}$ and $\mathrm{V_{max}}$ are, respectively, the minimum and maximum velocity determined based on the type of UAV and its mission. The UAV's acceleration is upper-bounded by $\mathrm{a_{max}}$, and $\mathbf{q}(t) \in \mathbf{q_{tr}}$ represents the instantaneous 3D location of UAV. 
For the UAV's trajectory, we adopt a random direction (RD) pattern following 3GPP study case \cite{3GPP36777}. According to RD pattern, the UAV starts its mission at a random location in 3D space and selects a random direction uniformly. Then, the UAV flies a given distance in a straight line with a speed bonded by UAV's application, capability, and the movement algorithm. 

\subsection{Propagation Channel} \label{subsec:channel}
We consider a dual-slop LoS/NLoS propagation channels for each link. Each propagation channel comprises 3GPP-based path-loss, small-scale fading, and 3D BSs antenna gain as described in the sequel.

\subsubsection{Path-Loss}
 The LoS and NLoS path-losses between UAV and k-\textit{th} BS can be respectively written as 
$
\mathrm{PL^\mathrm{LoS}_\mathrm{u,k}} = 28+22\log_{10}(\mathrm{d_\mathrm{u,k}})+20\log_{10}(\mathrm{f_c}),
$
and
$
\mathrm{PL^\mathrm{NLoS}_\mathrm{u,k}} = -17.5+[46-7\log_{10}(\mathrm{h_u})]\log_{10}(\mathrm{d_\mathrm{u,k}})  +20\log_{10}\left(\frac{40\pi\mathrm{f_c}}{3}\right),
$
where $\mathrm{d_\mathrm{u,k}}$ is the 3D distance between the UAV and the k-\textit{th} BS in meters, and $\mathrm{f_c}$ is the working frequency in GHz \cite{3GPP36777}. Note that path loss expressions are valid for $22.5\,\mathrm{m}<\mathrm{h_u}<300\,\mathrm{m}$ which is the range of interest for cellular-connected UAVs operations.

\subsubsection{Small-Scale Fading}
A wireless link between the UAV and \textit{k}-th BS undergoes small-scale fading with $\Omega_\mathrm{u,k}^\mathrm{LoS}$ and $\Omega_\mathrm{u,k}^\mathrm{NLoS}$ being the fading powers of LoS and NLoS conditions, respectively.
Without loss of generality, we adopt the convention $\mathbb{E}(\Omega_\mathrm{u,k}^\xi) = 1$, with $\xi$ being LoS or NLoS. We use the Nakagami-m fading model that covers a wide range of fading environments.  \cite{azari2019cellular}. Accordingly, $\Omega_\mathrm{u,k}^\xi$ follows a Gamma distribution, whose cumulative distribution function (CDF) can be found in \cite{azari2019cellular}.

\subsubsection{Probability of LoS/NLoS}
The aforementioned LoS and NLoS path-loss and small-scale fading components are incorporated to the system along with their probability of occurrence. For $22.5\,\mathrm{m}<\mathrm{h_u}<100\,\mathrm{m}$, the probability of LoS can be written as \cite{3GPP36777}
\[   
\mathrm{Pr^{LoS}_\mathrm{u,k}} = 
     \begin{cases}
       1, &\quad\text{if r}_\text{{u,k}}\le \mathrm{r_1}, \\
       \frac{\mathrm{r_1}}{\text{r}_\text{u,k}} + \left(1-\frac{\mathrm{r_1}}{\text{r}_\text{u,k}}\right) e^{\frac{-\text{r}_\text{u,k}}{\mathrm{r_2}}} , &\quad \text{if r}_\text{{u,k}} > \mathrm{r_1} \\ 
     \end{cases}
\]
where $\mathrm{r_1} = \max(460\log_{10}(\mathrm{h_u})-700,18)$ and $
\mathrm{r_2} = 4300\log_{10}(\mathrm{h_u})-3800$. Furthermore, for $\mathrm{h_u} \ge 100\,\mathrm{m}$ we have $\mathrm{Pr^{LoS}_\mathrm{u,k}} = 1$. Finally, the NLoS probability is  $\mathrm{Pr^{NLoS}_\mathrm{u,k}} = 1-\mathrm{Pr^{LoS}_\mathrm{u,k}}$.

\subsubsection{Antenna Gain}

We assume that each BS is equipped with a vertical N-element uniform linear array (ULA) with antenna element spacing of $0.5\lambda$. Each element has directivity of $\mathrm{A_E}(\theta,\phi)$, where $\theta$ and $\varphi$ are the spherical angles in local coordinate system of the origin at the antenna location. Following \cite{3GPP36873} the element gain can be written as
 $
     \mathrm{A_E}(\theta,\phi) = -\min\big\{-[\mathrm{A_{E,V}}(\theta)+\mathrm{A_{E,H}}(\phi)],\mathrm{A_{m}}\big\},
$
where 
$
    \mathrm{A_{E,V}}(\theta) = -\min\left\{12\left(\frac{\theta-90^o}{\theta_\mathrm{3dB}}\right)^2,\mathrm{SLA_v}\right\},
$
with $\theta_\mathrm{3dB} = 65^o,~\mathrm{SLA_v} = 30\,\mathrm{dB}$, and
$
    \mathrm{A_{E,H}}(\theta) = -\min\left\{12\left(\frac{\phi}{\phi_\mathrm{3dB}}\right)^2,\mathrm{A_m}\right\},
$
with $\phi_\mathrm{3dB} = 65^o,~\mathrm{A_m} = 30\,\mathrm{dB}$.
The maximum directional gain of an antenna element is considered to be $\mathrm{G_E^{max}} = 8$ dBi.

The total BS radiation pattern gain denoted by $G_b(\theta,\phi)$ is obtained as the superposition of each element's gain, i.e. $G_E(\theta,\phi) = \mathrm{G_E^{max}} + \mathrm{A_E}(\theta,\phi)$, and the array factor is given by 
$
\mathrm{G_A}(\theta) = 10\log_{10} \frac{\sin^2\Big(N\pi (\cos\theta - \cos\mathrm{\theta_t})/2\Big)}{N^2 \sin^2\Big(\pi (\cos\theta - \cos\mathrm{\theta_t})/2\Big)},	
$
where $\mathrm{\theta_t}$ is the electrical vertical steering angle defined between $0^o$ and $180^o$ ($90^o$ represents perpendicular to the array). Accordingly, $G_b(\theta,\phi) = G_E(\theta,\phi)+\mathrm{G_A}(\theta)$.

We also assume that the UAV is equipped with a single omnidirectional antenna of unitary gain in any direction. Therefore, the received power (in dB) at the UAV from k-\textit{th} BS can be expressed as
$
    P_k^\xi = 10\log_{10}\left(\frac{\mathrm{n_u}}{\mathrm{n_t}}\right) + \mathrm{P_b} + G_{b}(\theta_k,\phi_k) - \mathrm{PL}_\mathrm{u,k}^\xi + 10\log_{10}\left(\Omega^\xi_\mathrm{u,k}\right);~~\xi \in \{\mathrm{LoS},\mathrm{NLoS}\},
$
where $\theta_k$ and $\phi_k$ represent the spherical angles corresponding to the link from the k-\textit{th} BS to the UAV in 3D space. 


\subsection{Performance Metrics}\label{subsec:KPI}
In the following, we present important metrics that capture key limiting factors of mobile cellular-connected UAVs' performance.
\subsubsection{Disconnectivity}
During UAV's flight over $\mathbf{q_{tr}}$, the disconnectivity time corresponds to the amount of time that the serving BS is not able to provide a target rate $\mathrm{r_t}$ to the UAV. If we denote the UAV's achievable rate as $\mathcal{R}_\mathrm{D}$ then the link is called \textit{disconnected} if $\mathcal{R}_\mathrm{D} < \mathrm{r_t}$. 
The UAV's achievable rate $\mathcal{R}_\mathrm{D}$ is obtained as
$
    \mathcal{R}_\mathrm{D} = \mathrm{W}\,\mathbb{E}\big(\log_2(1+\mathsf{SINR})\big),
$
where W is the bandwidth assigned to the UAV's link which directly relates to the number of allocated PRBs, and $\mathsf{SINR}$ is the instantaneous signal-to-interference-plus-noise-ratio. If we assume that the J-\textit{th} BS is serving the UAV, the SINR at the UAV can be written as
$
    \mathsf{SINR} = \frac{P_J}{\sigma^2 + \sum_{k \ne J} P_k}
$
where $\sigma^2$ is the noise power.


\subsubsection{Handover Rate}
A UAV is able to constantly measure the received signal power from various BSs and eventually may decide to perform a new association based on the received power strengths. This HO procedure can significantly enhance the SINR level when switching to the best BS that provides the highest signal power. Since the HO procedure requires additional signaling exchanges in the network, it  yields  delays in communicating useful signals.
Although switching to the best BS increases the level of SINR and hence reduces the disconnectivity time, the UAV may encounter several consecutive HOs which, in turn, can result in a degraded reliability of the communication link. 


\subsubsection{Power Consumption}
One of the limiting factors in UAV's performance is their limited energy budget available for accomplishing a given mission. A UAV consumes energy for two major purposes including communication and propulsion. The energy consumption of the former particularly for C\&C  is negligible compared to the latter \cite{zeng2019accessing,sallouha2018energy}, hence we ignore the communication-related part. Assuming that the UAV maneuvering takes a small portion of the total operation time the power consumption of a rotary-wing UAV can be written as 
$
    P_{\mathrm{u}}(t) = P_0\left(1+\frac{3V^2}{U_\mathrm{tip}^2}\right) + P_i \left(\sqrt{1+\frac{V^4}{4\nu_0^2}}-\frac{V^2}{2\nu_0^2}\right)^{1/2}\!\!\!\!\!\!\! + K_\mathrm{u} V^3,
$
where $P_0$ and $P_i$ are constants respectively related to the blade profile power and induced hovering power, $U_\mathrm{tip}$ denotes the rotor blade's tip speed, $\nu_0$ represents the mean rotor induced hovering velocity, and $K_\mathrm{u}$ is determined by the fuselage drag ratio, rotor solidity, the air density, and rotor disc area. A more detailed discussion on the power consumption modeling of UAVs can be found in \cite{zeng2019accessing,sallouha2018energy}.  



\section{Preliminaries and Problem Statement} \label{sec:problem_statement}

This section considers critical aspects for the performance of cellular-connected UAV, and presents our problem statement. 

\subsection{Preliminaries}
\begin{figure}
\centering
\begin{subfigure}{\columnwidth}
\centering
\includegraphics[width=0.9\columnwidth]{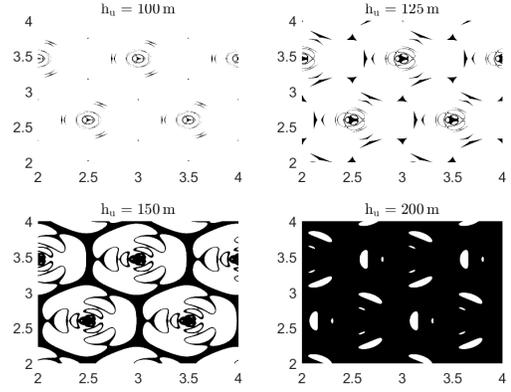}
	\caption{Black areas show disconnected areas where a target rate of 100\,kbps over one assigned PRB can not be satisfied.}
	\label{CoverageHoles_hUs}
\end{subfigure}
\newline
\begin{subfigure}{\columnwidth}
\centering
\includegraphics[width=0.9\columnwidth]{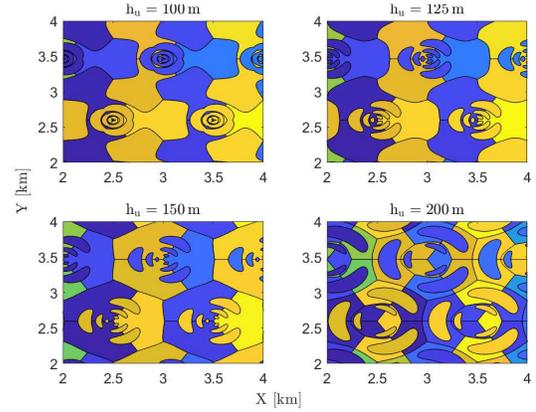}
	\caption{Crossing the lines trigger HO events. In general, the lines are denser at higher altitudes resulting in more HOs.}
	\label{HOlines_hUs}
\end{subfigure}
\caption{Limits illustration of cellular connectivity in the sky.}
\vspace{-5mm}
\end{figure}

Figure \ref{CoverageHoles_hUs} illustrates 
that disconnected areas (i.e. coverage holes) grow as the UAV's altitude increases resulting in a more disconnectivity time. This is due to the fact that the UAV at higher altitude experiences a strong LoS interference from the neighboring cells. 
Importantly, it may be impossible for a trajectory to avoid all the disconnected areas. However, the UAV could choose to pass over disconnected areas with a higher speed in order to reduce the time it remains disconnected from the network. Please note that allocating more PRBs to the UAV link results on a reduction of the disconnected areas. 
In this paper, however, our focus is to adopt proper UAVs speeds assuming that other cellular network-dependent parameters are fixed. Accordingly, no additional signaling overhead is imposed on the network.

To gain insight in the behaviour of the HO events, we have illustrated the cell association pattern at different altitudes in Figure \ref{HOlines_hUs},  where areas with the same color are served by the same cell. 
The black lines show the borders between any two different cells, and hence crossing any of these lines triggers a HO event. Importantly, the serving pattern greatly depends on the altitude: at higher altitudes the HOs lines tend to be denser, which implies that HOs happen more often.



\subsection{Problem Statement} \label{problem_statement}

We consider the perspective of a cellular-connected UAV designer who aims to accomplish a mission such as environmental sensing. For this, we consider a general objective that may include one or more of the followings: shortening the time of mission accomplishment, improving the life-time of UAV by reducing the energy consumption, reducing the disconnectivity time by considering a required target rate, or enhancing the link reliability by reducing the HOs rate. 

These PIs are linked since they are dependent on the UAV's velocity and the network topology. As the latter can not be controlled by a UAV operator, the former may be intelligently controlled to meet the objective's conditions. As the speed of the UAV increases, the time of mission accomplishment and disconnectivity time decrease. However, it yields an increased HO rate and it may increase or decrease the energy consumption\cite{zeng2019accessing,sallouha2018energy}. 
Accordingly, optimizing the UAV's speed is crucial to effectively balance the trade-offs between the time of mission accomplishment, UAV's lifetime, the connectivity time, and the reliability conditions. 
Mathematically, the problem can be formulated as: 
\begin{equation} 
\begin{aligned} 
\min _{V}  ~~ & f(T_{\mathrm{tr}},P_{\mathrm{u}},T_{\mathrm{D}},\mathcal{R}_{\mathrm{H}}) \\ \label{constraint_acceleration}
&\text { s.t. }  \mathrm{V_{\min }} \leq V \leq \mathrm{V_{\max }},~  |\dot{V}| \leq \mathrm{a_{max}}, ~
  \mathbf{q} \in  \mathbf{q}_\mathrm{tr},
\end{aligned}
\end{equation}
where $f(\cdot)$ represents a desired function which can be a \textit{weighted} summation 
of the arguments, $T_\mathrm{tr}$ is the time of task completion, $P_{\mathrm{u}}$ is the UAV's power consumption, $T_{\mathrm{D}}$ is the disconnectivity time and $\mathcal{R}_{\mathrm{H}}$ is the HO rate. Above, the first constraint represents the range of possible choices for velocity. The parameter $\mathrm{V_{\max} }$ is limited not only by the UAV's maximum possible speed but also by the type of task to be performed. 
Moreover, the limited acceleration of UAV is taken into account through \eqref{constraint_acceleration}. Note that the overall performance may not involve one or more of the considered PIs depending on the applications, and hence one may set some of the weights zero to focus on specific PIs. For instance, if a network can guarantee the required rate constraint for connectivity, then the disconnectivity weight in the objective function is set to zero. 





As the dependency of the objective function on the velocity is complex and determined by the environment which might not be known in advance, we propose a learning approach to solve the problem. 



\section{Learning-Based Controlled Mobility} \label{sec:ML_Solution}


Here we propose an RL based mechanism to address the problem stated in Eq. \eqref{constraint_acceleration}.

\subsection{An Overview of RL and MAB}

Reinforcement learning deals with the problem of designing policies for agents that need to act in environments whose inner workings they largely ignore. For building this policy, RL assumes that the agent has access to a reward signal, which provides feedback on how well the objective of the agent has been satisfied. The reward signal is updated after each action, which allows the agent to improve its policy based on its own past experiences. Importantly, RL does not rely on prior knowledge of the environment, being more flexible than other related frameworks such as supervised learning.

RL is a natural choice to address problems such as Eq. \eqref{constraint_acceleration}. Within the RL literature, our approach is to consider the multi-armed bandit (MAB), as it is one of the most well-understood scenarios with a vast related literature. A MAB abstracts a scenario where a gambler has to choose between a number of slot machines to play with. Each machine has its own likelihood of providing a positive outcome, which is not known by the gambler beforehand. Therefore, the gambler needs to do exploratory rounds to estimate the odds of each machine, to then exploit this knowledge by playing in the most favorable ones. Importantly, each exploratory round improves the knowledge of the gambler about the machines at the expense of risking low payoffs; nevertheless, this knowledge is key for ensuring favorable long-term outcomes. This exploration-vs-exploitation trade-off is a hallmark of RL problems, and its balance is a main concern of RL algorithms.

\subsection{Mobility Management as a MAB Problem}

To solve \eqref{constraint_acceleration} as a MAB problem, the UAV acts as the agent with the set of available velocities as the actions $\mathcal V = \{\mathrm{V_{min}}, \mathrm{V_{min}}+\Delta V,  \dots, \mathrm{V_{max}}\}$. The UAV's path is divided into \textit{m} equal segments. At the beginning of each segment the UAV selects a velocity from $\mathcal V$ -- within the ones that are consistent with the maximal acceleration, i.e. $\mathrm{a_{max}}$. Then, the UAV computes a reward signal defined as
\begin{align} \label{reward}
    r_\mathrm{v}(t) = &\Big(\beta_\mathrm{V}+\beta_\mathrm{D} \, \rho_\mathrm{D}(t)- \beta_\mathrm{H} \, \bar{n}_\mathrm{H}(t) \Big)\frac{V(t)-V_\mathrm{min}}{V_\mathrm{max}-V_\mathrm{min}} \nonumber \\ &- \beta_\mathrm{P} \, \frac{P_u(t)-P_\mathrm{min}}{P_\mathrm{max}-P_\mathrm{min}},
\end{align}
where the coefficients $\beta_\mathrm{V}$, $\beta_\mathrm{D}$, $\beta_\mathrm{H}$, and $\beta_\mathrm{P}$ are learning parameters that indicate the impact of mission completion time, disconnectivity time, HOs, and power consumption on the reward function, respectively. Since increasing the UAV's speed decreases the traveling time, it is considered as a benefit in the reward function. In \eqref{reward}, $\rho_\mathrm{D}$ denotes the fraction of the time interval at segment \textit{t} when the UAV is in the disconnectivity condition. 
Note that the time duration of disconnectivity can be decreased by increasing the speed. In \eqref{reward}, moreover, $\bar{n}_\mathrm{H}$ is the average number of HOs occurred until the segment \textit{t} which is considered as a cost. Therefore, the  cost due to HOs increases with a higher number of HOs. The last term in \eqref{reward} captures the impact of power consumption on the reward function as a cost. Finally, the speed and power are normalized within the range [0,1] in order to fairly combine the benefits and costs elements.

\subsection{Solution Based on Upper Confidence Bound (UCB)}
The upper confidence bound (UCB) algorithm is an effective approach for solving the MAB problem. Using the UCB algorithm, the UAV  first selects each velocity once. Then, as the iteration becomes larger than the number of actions denoted as $N$, it selects the velocity $V_{\mathrm{UCB}}^*(t)$ according to the decision function that satisfies
\begin{equation} \label{UCB}
V_{\mathrm{UCB}}^*(t) = \mathop{\argmax }_{\mathrm{v} \in \tilde{\mathcal V}} \left\{\Bar{r}_\mathrm{v}(t)+ c \sqrt{\frac{2 \ln t}{n_\mathrm{v}(t)}}\right\}
\end{equation}
where $\Bar{r}_\mathrm{v}(t)$ denotes the mean reward of velocity $\mathrm{v}$ until segment $t$, $n_\mathrm{v}(t)$ is the number of times that arm $\mathrm{v}$ has been selected, and $\tilde{\mathcal V}$ is a subset of $\mathcal V$ ensuring the acceleration constraint in \eqref{constraint_acceleration}. In  \eqref{UCB}, the mean reward term and  the second term capturing $n_\mathrm{v}(t)$ correspond to exploitation and exploration, respectively.  Parameter $c>0$ balances the trade-off between exploration and exploitation. A pseudocode describing our proposed algorithm is presented in Algorithm \ref{alg_ucb}.

\begin{algorithm}[h]
\scriptsize
\caption{:   UCB-based   mobility management algorithm for the cellular-connected UAV}
\label{alg_ucb}
\begin{algorithmic}[1]
\STATE {\textit{{Initialization:}}}
\STATE $\bar{n}_\mathrm{HO}(0) = 0$
\FOR{$t = 1:N$} 
\STATE Select the t-\textit{th} arm $\mathrm{V}\in \mathcal V$
\STATE measure $\rho_\mathrm{D}(t)$, $n_\mathrm{HO}(t)$, and $P_\mathrm{u}(t)$
\STATE Assign $\bar{n}_\mathrm{HO}(t) = \big[(t-1)\bar{n}_\mathrm{HO}(t-1)+n_\mathrm{HO}(t)\big]/t$
\STATE Calculate $r_\mathrm{v}(t)$ using \eqref{reward} 
\STATE Assign $\Bar{r}_\mathrm{v}(t) = r_\mathrm{v}(t)$ 
\STATE Assign $n_\mathrm{v}(t) =1$

\ENDFOR
\STATE {\textit{Main loop:}}
\FOR{$t>N$}
\STATE Select arm $V_{\mathrm{UCB}}^*(t) $ according to \eqref{UCB}
\STATE measure $\rho_\mathrm{D}(t)$, $n_\mathrm{HO}(t)$, and $P_\mathrm{u}(t)$
\STATE Assign $\bar{n}_\mathrm{HO}(t) = \big[(t-1)\bar{n}_\mathrm{HO}(t-1)+n_\mathrm{HO}(t)\big]/t$
\STATE Calculate $r_\mathrm{v}(t)$ according to \eqref{reward}
\FOR{$\forall \mathrm{v}\in \mathcal V$}
\STATE Update $n_\mathrm{v}(t) = n_\mathrm{v}(t-1) + \mathds{1}_{\{\mathrm{v} = \mathrm{v}_{\mathrm{UCB}}^*(t)\}}$
 \STATE Update $\Bar{r}_{\mathrm{v}}(t) = \big[n_\mathrm{v}(t-1) \Bar{r}_{\mathrm{v}}(t-1)+\mathds{1}_{\{\mathrm{v} = \mathrm{v}_{\mathrm{UCB}}^*(t)\}}r_\mathrm{v}(t)\big]/n_\mathrm{v}(t)$

\ENDFOR
 \STATE $t\leftarrow t+1$
\ENDFOR

\end{algorithmic}
\end{algorithm}%

\section{Numerical Results} \label{sec:numerical_results}

\begin{figure}[!t]
\centering
\includegraphics[width=0.9\columnwidth]{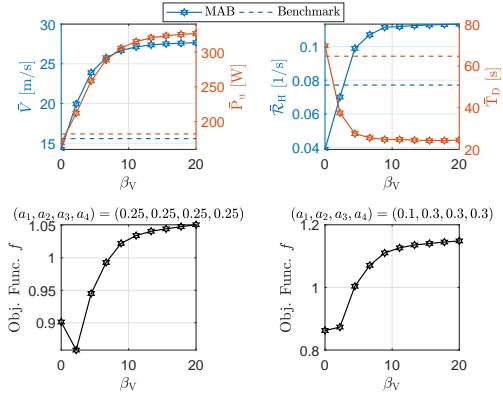}
	\caption{Considering the overall performance indicator, i.e. $f$, learning strategy outperforms benchmark method for low values of $\beta_\mathrm{V}$.}
	\label{metrics_betaV}
	\vspace{-3mm}
\end{figure}

\begin{figure}[!t]
\centering
\includegraphics[width=0.9\columnwidth]{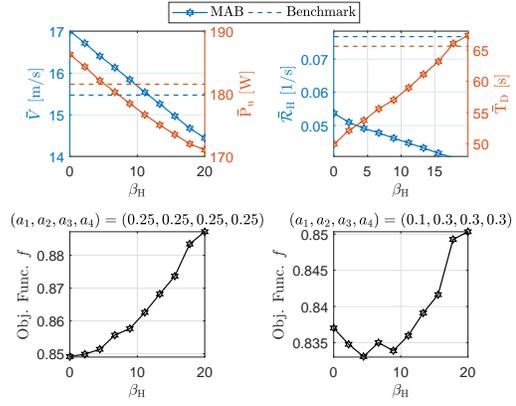}
	\caption{HO cost can be reduced by increasing $\beta_\mathrm{H}$. However, the overall performance is optimized within a limited range of $\beta_\mathrm{H}$.}
	\label{metrics_betaH}
	\vspace{-3mm}
\end{figure}

In this section, we examine our proposed algorithm by focusing on each target performance individually and also in total on the objective function. To study the total effect of the parameters $\beta_\mathrm{V},\beta_\mathrm{D},\beta_\mathrm{H},\beta_\mathrm{P}$, we consider an objective function given by
 $   f(T_{\mathrm{tr}},P_{\mathrm{u}},T_{\mathrm{D}},\mathcal{R}_{\mathrm{H}}) = a_1\, \bar{T}_{\mathrm{tr}}/\bar{T}_{\mathrm{tr}}^\mathrm{B} + a_2\, \bar{P}_{\mathrm{u}}/\bar{P}_{\mathrm{u}}^\mathrm{B} + a_3\, \bar{T}_{\mathrm{D}}/\bar{T}_{\mathrm{D}}^\mathrm{B} + a_4\, \bar{\mathcal{R}}_{\mathrm{H}}/\bar{\mathcal{R}}_{\mathrm{H}}^\mathrm{B},
$
where superscript B denotes the benchmark result, $\bar{\cdot}$ indicates the average, and $a_i$s (i=1,2,3,4) are non-negative real coefficients where $\sum_{i=1}^{4} a_i = 1$. This representation of $f$ enables us: 1) to determine the importance of each individual metric as compared to others by adjusting $a_i$s (For an specific scenario some of the $a_i$s may be set equal to zero), 2) to fairly evaluate the overall impact of the learning method as compared to the benchmark. Please note that, $f$ is equal to 1 for the benchmark and improvements due to learning should be reflected in performances where $f<1$. 
The benchmark is obtained by adopting a uniform random velocity in each iteration. The parameters used for the simulations are summarized in Table \ref{tab:parameters}. The default values of the learning parameters are one. Furthermore, for the parameters' values of power consumption and BSs antenna pattern, see \cite{zeng2019accessing} and \cite{3GPP36873} respectively.

\begin{table}[h!] 
	\centering
	\caption{Notations and values for simulation.}
	\begin{tabular}{|c|c|c|c|c|c|c|}
		\hline\hline
		 $\mathrm{N_0}$ & $\mathrm{r_t}$ & $\mathrm{W}$ & $\mathrm{f_c}$ \\ \hline 
		 noise power & target rate & bandwidth & carrier frequency  \\ \hline
		 -204 dB/Hz & 100\,kbps & 180\,kHz & 2\,GHz \\
		\hline\hline
		 UMa  & ISD & $\mathrm{h_b}$ & $\mathrm{P_b}$ \\ \hline
		environment & inter-site distance & BSs height & BSs transmit power \\ \hline
		 6\,km $\times$ 6\,km & 1\,km & 25\,m & 46\,dBm \\ \hline \hline
		 $\mathrm{V_{min}}$  & $\mathrm{V_{max}}$ & $\mathrm{a_{max}}$ & $\mathrm{h_u}$ \\ \hline
		min. velocity & max. velocity & max. accel. & flying altitude \\ \hline
		 1\,m/s & 30\,m/s & 5\,m/s$^2$ & 150\,m \\ \hline \hline
	\end{tabular} \label{tab:parameters}
\end{table}

Figure \ref{metrics_betaV} illustrates the impact of velocity learning parameter $\beta_\mathrm{V}$ on each PI and also on the total objective function $f$. As expected, an increase in $\beta_\mathrm{V}$ increases the velocity, and hence proportionally reduces the time of task completion. A higher $\beta_\mathrm{V}$, also, decreases the disconnectivity time $\mathrm{T_D}$, however, it is disadvantageous for handover rate and power consumption. As can be seen, the total effect of $\beta_\mathrm{V}$ on the objective function can be minimized by adopting proper values of $\beta_\mathrm{V}$. For instance for the case of $a_i = 0.25,~i=1,2,3,4$, the best choice is $\beta_\mathrm{V} \approx 2 $.

Figure \ref{metrics_betaH} shows the higher HO cost $\beta_\mathrm{H}$ results in a  lower HO rate and velocity. The HO rate is significantly lower than the benchmark using the learning technique. Altogether, the total effect of HO learning parameter can be balanced by choosing $\beta_\mathrm{H}$ between 0 and 5 in our examined cases. It is worth pointing out that in these figures for $\beta_\mathrm{H} \in [0,20]$ still the learning method outperforms the benchmark as the objective function is below 1.

\begin{figure}[!t]
\centering
\includegraphics[width=0.9\columnwidth]{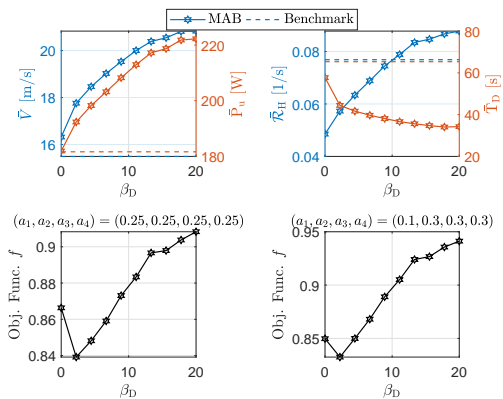}
	\caption{Increasing $\beta_\mathrm{D}$, i.e. the disconnectivity learning parameter, reduces the disconnectivity time and increases the HO rate. Overall, there is an optimal value of $\beta_\mathrm{D}$ that balances all the effects in the objective function.}
	\label{metrics_betaD}
	\vspace{-3mm}
\end{figure}

Figure \ref{metrics_betaD} shows that an increase in the disconnectivity learning factor $\beta_\mathrm{D}$ reduces the disconnectivity time which confirms the suitability of the reward function in \eqref{reward} for this problem. Furthermore, with increasing $\beta_\mathrm{D}$, power consumption increases. The main reason is that the UAV speed increases compared to the optimal speed for  the minimum power consumption. 
The overall performance represented by the objective function can be minimized by choosing $\beta_\mathrm{D} \approx 2$, though for the other examined values yet the learning approach benefits the system as $f$ lies below 1. 

\begin{figure}[!t]
\centering
\includegraphics[width=0.9\columnwidth]{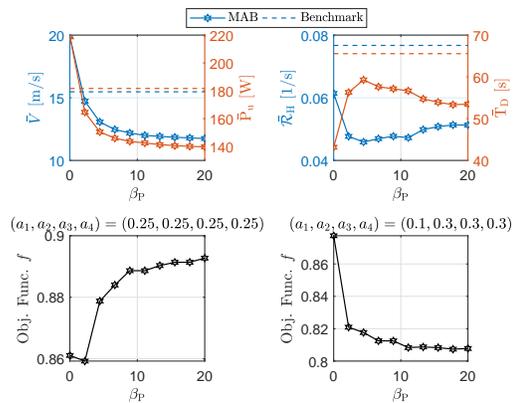}
	\caption{Depending on the weight of each performance indicator in the objective function, increasing power consumption's learning parameter, i.e. $\beta_\mathrm{P}$, can be detrimental (left figure) or beneficial (right figure).}
	\label{metrics_betaP}
	\vspace{-3mm}
\end{figure}

Figure \ref{metrics_betaP} reveals that the higher values of power consumption cost $\beta_\mathrm{P}$ makes this contributor dominant over others, and hence the velocity converges to the optimum velocity for the minimum power consumption. This fact results in a relatively stable behavior of HO rate and disconnectivity time for large values of $\beta_\mathrm{P}$. In general, the average HO rate and disconnectivity time are not monotonic functions of $\beta_\mathrm{P}$ motivated by the fact that the dependency of power consumption on velocity is not monotonic. As for the objective function $f$, one can see that depending on the weight of each term, i.e. $a_i$, the larger or lower values of $\beta_\mathrm{P}$ is better. Reducing $a_1$, i.e. the importance of time completion, and increasing the weight of power consumption and other individual metrics require higher values of $\beta_\mathrm{P}$ for an optimal overall performance.

\section{Conclusion} \label{sec:conclusion}

In this paper, we have addressed key challenges of mobile cellular-connected UAVs including connectivity time, handover rate, energy consumption, and traveling time by using a reinforcement learning approach. Our approach leverages the Upper Confidence Bound algorithm for MAB problems, which we recast in the context of cellular-connected UAV systems by recognition of a proper reward function. 
Our results show that adequate learning parameters enable significant improvement in the key performance indicators. Interestingly, the optimal combination of learning parameters depends on the weight of each indicator.

\bibliographystyle{IEEEtran}
\bibliography{Bib_Mahdi}

\end{document}